\documentclass[preprint,showpacs,preprintnumbers,amsmath,amssymb]{revtex4}
\usepackage{graphicx}
\usepackage{dcolumn}
\usepackage{bm}
\usepackage[usenames]{color}
\newcommand{\be}{\begin{equation}}
\newcommand{\en}{\end{equation}}
\newcommand{\bea}{\begin{eqnarray}}
\newcommand{\ena}{\end{eqnarray}}

\begin{document}


\title{ Intermediate inflation into generalized induced-gravity scenario }

\author{Carlos Gonz\'alez}
\email{carlos.gonzalez@uda.cl} \affiliation{ Departamento de F\'{\i}sica, Universidad de Atacama, Avenida Copayapu 485, Copiap\'{o}, Chile.}

\author{Ram\'on Herrera}

\email{ramon.herrera@pucv.cl} \affiliation{ Instituto de
F\'{\i}sica, Pontificia Universidad Cat\'{o}lica de
Valpara\'{\i}so, Avenida Brasil 2950, Casilla 4059,
Valpara\'{\i}so, Chile.}

\date{\today}
\begin{abstract}

An intermediate inflationary universe model within the context of
non-minimally coupled to the scalar curvature is analyzed. We will conduct our analysis
under the slow roll approximation of the inflationary dynamics
and the cosmological perturbations considering a coupling of the
form $F(\varphi)=\kappa+\xi_n\varphi^n$. Considering the
trajectories in the $r-n_s$ plane from Planck data, we find the
constraints on the parameter-space in our model.
\end{abstract}

\maketitle

\section{Introduction}
It is well known that the inflationary era, or simply inflation,  has been a
fundamental contributor to our understanding of the early Universe. In this sense,
 inflation has been successful in explaining
some of cosmological puzzles, i.e., the horizon and  flatness
problems etc. \cite{Sta,guth, linde1983}. However, the most
important
 element in  this  scenario,  is that inflation
 gives us a  theoretical framework within which
to describe   the large-scale structure (LSS)\cite{est}, as well as  the
 anisotropy of the cosmic microwave background (CMB)
radiation from the early Universe\cite{CMB}.  From  observational point of view, the Planck satellite \cite{Pl2013}
 together with the LSS
experiments \cite{New1} (in particular considering  the Baryon Acoustic Oscillations (BAO) data), have been
 fundamental  to our understanding of the CMB anisotropies of the Universe.

With respect to the inflationary  epoch, it  thought that it is
driven by a scalar field or simply inflaton  that  can interact
fundamentally  with other fields, and also with the
gravity.  In this sense, it is natural to consider
 the
non-minimal coupling between the scalar field and the
gravitation. Historically, the non-minimal coupling with gravity and in particular with  the   Ricci scalar, was originally studied in radiation problems
\cite{rad}. Within the context of
 the renormalization of the quantum
fields in curved backgrounds was considered in
 Refs.\cite{r1,r2}. From the point of view of cosmology this non-minimal
 coupling of the scalar field was originally analyzed  in Ref.\cite{Jor},
 and also for the authors
 Brans and Dicke \cite{BD}, see also
 Ref.\cite{B1}. In the literature of the  Eighties, the cosmic inflation in the framework of the induced gravity
 scalar-tensor theory has been studied in Refs.\cite{81,82,83}.
  In particular, a description of the inflationary models
 considering
the non-minimal coupling between the scalar field and the
gravitation, has been developed in  Refs.\cite{Spo,val,fakir,
faraoni,Amendola:1999qq,Uzan:1999ch}. Particularly, in
Ref.\cite{fakir} was studied  the chaotic model considering  this
coupling. In Ref.\cite{futamase} was assumed  the effective
chaotic potential $V\approx\varphi^{n}$ in which $n>4$,
considering one large scale for the field  $\varphi$ within the
framework to this coupling. Here, the authors obtained  different
constraints on the parameter-coupling $\xi$, also see
 Ref.\cite{Bb2}. For the relation between the tensor to scalar ratio and the scalar spectral index
 i.e., the consistency relation was
studied in Ref.\cite{new1} and applied to chaotic inflation model,
 and  a global stability analysis also was studied in
Ref.\cite{new2}. Recently, the frame-independent classification of
single-field inflationary models was considered in
Ref.\cite{Jarv:2016sow}, and the important case of Higgs inflation
in the non-minimally coupled inflation sector was studied in Ref.\cite{bezrukov}.

Within the context of the dynamics background for the inflationary model,   exact solutions
within the framework of the General Relativity (GR)
can be found when the scale factor expands exponentially from  a constant potential,  ``de  Sitter''
inflationary model\cite{guth}. Similarly, an expansion of the power-law type, in
which the scale factor $a(t)\propto t^p$ with $p>1$, can  be obtained from an exponential potential  giving
exact solutions\cite{power}. Nevertheless, intermediate inflation  is
other kind of exact solution, in which  the expansion of the scale factor
 is slower than de Sitter expansion, but
faster than power-law type.  During intermediate expansion
the scalar factor $a(t)$ expands as
\begin{equation}
a(t)=\exp[A\,t^f],\label{int}
\end{equation}
where $A$ and $f$ are two constants, in which  $A>0$ and
$0<f<1$\cite{int}. As mentioned previously, this  inflationary
model was originally studied in order to find an exact solution to
the background equations. Nevertheless from the observational
point of view, intermediate inflation  is more effectively
motivated  under the slow-roll approximation \cite{Barr}. 
Here, considering
 the slow-roll approximation,
 it is feasible
to find a scalar spectral index $n_s\sim 1$, and this kind of
spectrum is favored by the current CMB data. In particular  for
the specific value of $f=2/3$ in which the scale factor varies  as
$a(t)\propto t^{2/3}$, the scalar spectral index becomes $n_s=1$,
i.e., the Harrizon-Zel'dovich spectrum. However, within the  GR
framework this model presents a fundamental problem due to the
fact that the tensor-to-scalar ratio $r>0.1$,  wherewith the
model is disfavored from observational data, as a result of the
Planck data's establishment of an upper bound for the ratio $r$ at
pivot scale $k_*=0.002$Mpc$^{-1}$, in which $r_{0.002}<0.1$.  In
this form, the model of intermediate inflation in the GR framework
does not work.


In this paper we would like to study the possible actualization of
an expanding intermediate inflation within the framework of a
non-minimal coupling with the curvature. We will explore the
dynamics from the slow roll approximation  in this theory
considering a function $F(\varphi)=\kappa+\xi_n\varphi^n$. Within this
context, we will
 find    the cosmological perturbations; scalar perturbation and tensor
 perturbation, and from the  trajectories in the $r-n_s$ plane, and  we will establish whether  or not the model works.
In our analysis, we shall resort to  Planck
satellite\cite{Ade:2013uln} in order to constrain the parameters
in our model.

The outline of the article is as follows. The next section
presents the background dynamics and we find  the slow roll
solutions for our model. In Sect. III we determine the
corresponding cosmological perturbations. Finally, in Sect. IV we
summarize our finding. We chose units so that $c=\hbar=1$.

\section{Intermediate inflation: Background equations }

We consider a generalized induced- gravity action in Jordan frame
given by
\begin{equation}
S=\int d^4x \sqrt{-g}\left[ \frac{F(\varphi)}{2} R
-\frac{1}{2}g^{\mu\nu}\varphi_{;\mu}\varphi_{;\nu} - V
(\varphi)\right], \label{eq1}
\end{equation}
where $F(\varphi)$ can be an arbitrary function of the scalar
field in the Jordan frame,  $R$ is the Ricci scalar and
$V$($\varphi$) is the effective potential associated to the scalar
field $\varphi$.

Different types of the functions $F(\varphi)$  have been studied in the
literature. In particular when the function $F(\varphi)=(1-\xi\varphi^2)$ coincides with the non 
minimal coupling action, see Ref.\cite{ab}. Also, the special case in
which the function $F(\varphi)\propto\varphi^2$ was studied in
\cite{kaiser}. In the following, we will assume that the function
$F(\varphi)$ is defined as
\begin{equation}
F(\varphi)=\kappa+\xi_n\varphi^{n},\label{eq10}
\end{equation}
where $\kappa=\frac{1}{8\pi G}=\frac{M_p^2}{8\pi}$, with $M_p$ is
the reduced Planck mass and  the quantities $n$ (dimensionless)
and $\xi_n$ (with units of $M_p^{2-n}$) are two constants. In the
particular case in which the exponent of the function corresponds
to  $n=2$ ($\xi_{n=2}=\xi$) together with $\kappa=0$, coincides
with that corresponding induced-gravity model studied in
Ref.\cite{accetta}.

From the action given by Eq.(\ref{eq1}), the cosmological
equations  in a spatially flat Friedmann-Robertson- Walker (FRW)
cosmological model in the Jordan frame are given by

\begin{equation}
H^2=\frac{1}{6F}\left[\dot{\varphi}^2+2V - 6H\dot F\right],\label{eq2}
\end{equation}

\begin{equation}
\dot{H}=\frac{1}{2F}\left[-\dot{\varphi}^2+H\dot F -\ddot F\right],\label{eq3}
\end{equation}
and
\begin{equation}
V_{,\,\varphi}= 3F_{,\,\varphi} (\dot{H}+2H^2)
-3H\dot{\varphi}-\ddot{\varphi},\label{eq4}
\end{equation}
where $H=\dot{a}/a$ is the Hubble parameter and  $a$ corresponds
to  the scale factor. Here, the dots means derivatives with
respect to time and the subscription $(,\,\varphi)$ means
derivatives with respect to scalar field $\varphi$.

Introducing the slow- roll parameters defined by

\begin{equation}
\epsilon_{1}\equiv
-\frac{\dot{H}}{H^{2}},\;\;\;\;\;\epsilon_{2}\equiv
\frac{\ddot{\varphi}}{H\dot{\varphi}},\;\;\;\;\epsilon_{3}\equiv
\frac{\dot{F}}{2HF},\;\;\;\;\mbox{and}\;\;\;\;\epsilon_{4}\equiv
\frac{\dot{E}}{2HE},\label{eq5}
\end{equation}
in which $\epsilon_{i}\ll1$ and its evolution
$\dot\epsilon_{i}\simeq 0$ during inflation. Here, the quantity
$E$ is defined as $E= F(1+\frac{3\dot{F}^2}{2F\dot{\varphi}^2})$.

Considering the slow roll parameter $\epsilon_{3}$, we have
$\frac{\ddot{F}}{2F}=H\dot \epsilon_{3}$, then we can neglect the
last right side term in Eq. (\ref{eq3}), thus during the slow roll
scenario we get

\begin{equation}
2F\dot H + \dot\varphi^{2}-H\dot F\simeq0.\label{eq9}
\end{equation}

Now combing Eqs.(\ref{int}), (\ref{eq10}) and (\ref{eq9}) we find
that the  differential equation for the scalar field is given by


\begin{equation}
2Af(f-1) t^{f-2}(\kappa+\xi_n\varphi^{n}) + \dot{\varphi}^{2} -
Afn\xi_n t^{f-1}\varphi^{n-1}\dot{\varphi}=0.\label{eq12}
\end{equation}

The  solution of Eq.(\ref{eq12}) for the scalar field  can be
written as

\begin{equation}
\varphi(t)=c\,t^{m},\label{eq13}
\end{equation}
where $m$ and $c$ are two  constants. In order to satisfy the
power law solution of the scalar field given by Eq.(\ref{eq13}),
we find the constraints
$$
m=\frac{f}{2},\;\;\;\;c^{2}=\frac{8\kappa A
(1-f)}{f},\,\,\,\,\,\mbox{and}\,\,\,\,\,\,\,n=4(1-\frac{1}{f}).
$$
As we have mentioned previously  the parameter $f$ lies between
$0<f<1$, in order to obtain an acceleration of the Universe, then
the parameter $n<0$.

By using the slow roll approximation on Eq.(\ref{eq2}) and
 combining Eqs.(\ref{int})  and
(\ref{eq13}),  the effective potential in terms of the scalar
field results

\begin{equation}
V(\varphi)\approx 3[H^2\,F+H\dot{F}]=3Afc^{-n}\left[ Af\kappa
 + Af\xi_n\varphi^{n} +  \xi_nnmc^2\varphi^{n-2}\right]\varphi^{n}.
\label{eq16}
\end{equation}
Here,  we note that considering the special case in which
$\xi_n=0$ (or equivalently $F=const$.), the effective potential
given by Eq.(\ref{eq16}) coincides with that corresponding to the
standard intermediate inflation in General Relativity, where
$V(\varphi)\propto\varphi^{4(1-f^{-1})}$ see Ref.\cite{Barr}.

An important quantity during the background dynamics corresponds
to the number of e-folds $N$ at the end of inflation and becomes
\begin{equation}
N = \int _{t_1}^{t_{2}} Hdt =
\int_{\varphi_1}^{\varphi_{2}}\frac{H(\varphi)}{\dot{\varphi}}d\varphi=
\frac{f}{8\kappa
(1-f)}\lbrace\varphi_{2}^{2}-\varphi_{1}^{2}\rbrace,\label{eq24}
\end{equation}
between two different values of cosmological times $t_1$ and $t_2$
or between $\varphi_1$ and $\varphi_2$ values of the scalar field.

 On the other hand, the parameter $\epsilon_1$ from
Eqs.(\ref{eq5}) and (\ref{eq13}) can be written as
\begin{equation}
\epsilon_1=\frac{8\kappa (1-f)^2}{f^2}\,\frac{1}{\varphi^2}.
\end{equation}
Here, we note that the slow roll parameter $\epsilon_1$ diverges
when the scalar field $\varphi\rightarrow 0$ and then
$\epsilon_1\gg 1$ (or equivalently $\ddot{a}\ll 0$, deceleration).
In this limit when the field approaches to zero,  the effective
potential  given by Eq.(\ref{eq16}) also diverges (since the
exponent $n<0$). Now, for  large values of the scalar field, we
note that  asymptotically the effective potential and the slow
roll parameter $\epsilon_1$ tend to zero, in which $\epsilon_1<1$.
In this sense, since during the evolution of the universe the
effective potential decreases, we consider that the inflation
scenario begins at the earliest possible epoch, in which
$\epsilon_1=1$, where  the scalar field $\varphi_1=\varphi(t_1)$
 results
\begin{equation}
\varphi_{1}=\sqrt{8\kappa}\frac{(1-f)}{f}.\label{eq26}
\end{equation}
Also, we note  that the inflationary epoch takes place when
$\epsilon_1<1$ (or equivalently $\ddot{a}>0$), then the scalar
field satisfies the condition $\varphi>\frac{\sqrt{8\kappa}
(1-f)}{f}$.

In this way, we find that the value of the scalar field
$\varphi_2=\varphi(t=t_2)$ in terms of the number of e-fold $N$
can be written as
\begin{equation}
\varphi_2^2=\frac{8\kappa (1-f)(1+f(N-1))}{f^{2}}.\label{eq27}
\end{equation}
Here, we have used Eqs.(\ref{eq24}) and (\ref{eq26}).

\section{Cosmological Perturbations}

In this section we will study the cosmological perturbations in
our model of intermediate inflation into generalized
induced-gravity scenario. In this framework, the  perturbation
metric around the flat background can be written as
\begin{equation}
ds^{2}=-(1+2\Phi)dt^2+2a(t)\Theta_{,i}dx^idt+a^2(t)[(1-2\psi)\delta_{ij}+2E_{,i,j}+2h_{ij}]dx^idx^j,\label{me}
\end{equation}
where the quantities $\Phi$, $\Theta$, $\psi$ and $E$ correspond
to the scalar-type metric perturbations, and  $h_{ij}$ denotes the
transverse traceless tensor-perturbation. Also the perturbation in
the  field $\varphi$ is given by $\varphi(t,\vec{x}
)=\varphi(t)+\delta\varphi(t,\vec{x} )$, in which
$\delta\varphi(t,\vec{x} )$  is  a small perturbation that
corresponds to  small fluctuations of the scalar field. Thus,
introducing the comoving curvature perturbations, ${\cal{R}}$
defined as ${\cal{R}}= \Psi +
\mathcal{H}\frac{\delta\varphi}{\varphi^{\prime}}$\cite{Bardeen},
where the new Hubble parameter is given by
$\mathcal{H}\equiv\frac{a^{\prime}}{a}$ (a prime corresponds to
the derivative with respect to a conformal time $d\eta=a^{-1}
dt$), then the  scalar density perturbation ${\cal{P}}_S$ in the
Jordan frame can be written as  \cite{noh,shinji}
\begin{equation}
{\cal{P}}_{S}\equiv \frac{k^{3}}{2\pi^{2}}|{\cal{R}}|^{2} =
A_S^{2}\left[\frac{k |\eta |}{2}\right]^{3-2\nu_{s}},\label{pert}
\end{equation}
where the amplitude $A_{S}^{2}\equiv
\frac{1}{Q_{s}}(\frac{H}{2\pi})^{2} (\frac{1}{aH|\eta|})^{2}
[\frac{\Gamma(\nu_{s})}{\Gamma(3/2)}]^{2} $ and the quantities
$Q_s$ and $\nu_{s}$ are defined as
$$
Q_s=\frac{\varphi^{\prime\,2}\left[1+\frac{3\xi_nn^2\varphi^{n-2}}{2}\right]}
{\left[\mathcal{H}+\frac{n\varphi^{\prime}}{2\varphi}\right]^2},\,\,\,\,\mbox{and}\,\,\,\,\nu_{s}
=\sqrt{\gamma_{s} + 1/4},
$$
respectively. Here, the parameter $\gamma_s$ is given by
$\gamma_{s} =\frac{(1+\delta_{s})(2-\epsilon_1 +
\delta_{s})}{(1-\epsilon_1)^{2}},$ in which
$\delta_{s}=\frac{\dot{Q_{s}}}{2HQ_{s}}$.

On the other hand, the scalar spectral index $n_s$ is defined a
$n_s=1+\frac{d\ln {\cal{P}}_{S}}{d\ln k}$. Thus, from
Eq.(\ref{pert}) and considering  the slow-roll approximations, the
scalar spectral index $n_s$ in terms of the slow roll parameters
can be written as \cite{noh}
\begin{equation}
n_{s}\simeq 1-2(2\epsilon_{1} + \epsilon_{2} - \epsilon_{3}+
\epsilon_{4}).\label{eq17}
\end{equation}

By considering the slow roll parameters from Eq.(\ref{eq5}), the
scalar spectral index $n_s$ as a function of the scalar field
i.e., $n_s=n_s(\varphi)$ results
\begin{equation}
n_{s}\simeq1-C_1\varphi^{-2}-C_2\varphi^{n-2}\left[\frac{1+3\xi_nn(n-1)\varphi^{n-2}}{\kappa+\xi_n\varphi^n+
\frac{3}{2}\xi_n^2n^2\varphi^{2(n-1)}}-\frac{1}{\kappa+\xi_n\varphi^{n}}\right],\label{eq19}
\end{equation}
where the constants $C_1$ and $C_2$ are defined as
$$
C_1=\frac{16\kappa(1-f)(1-\frac{3}{2}f)}{f^2},\,\,\,\,\,\,\mbox{and}\;\;\;\;C_2=\frac{4\xi_nn\kappa
(1-f)}{f}.
$$
From Eq.(\ref{eq19}) we observe that in the limit
$\xi_n\rightarrow 0 $ the spectral index $n_s$ agrees with the
appropriate standard intermediate inflationary model, in which
$n_s=1-C\varphi^{-2}$ with $C=8\kappa(1-f)(2-3f)/f^2$, see
Ref.\cite{Barr}. Also we noted that for the specific value $f=2/3$
we clearly find from Eq.(\ref{eq19}) that  $n_s\neq 1$, contrarily
to  standard intermediate inflation, where for $f=2/3$ the index
$n_s=1$, i.e., so-called Harrizon-Zel'dovish spectrum\cite{Barr}.
In this form, we noted from Eq.(\ref{eq19}) that the spectral
index includes  two free parameters $(f,\xi_n)$ unlike that occurs
in the standard intermediate inflation, where $n_s$ has only one
free-parameter, namely $f$. In relation to the constants,
 we note that for values of $f \leq 2/3$ the
constant $C_1 \geq 0$. Similarly, we observed that as the
parameter $n<0$, then the constant $C_2\geq 0$ when the
parameter $\xi_n\leq 0$.

 Also, we find that the spectral index  $n_s$ in terms of the
number of e-fold, $N$, becomes
\begin{equation}
n_{s}\simeq1-\frac{C_1}{\kappa\beta^2}-\frac{C_2}{\kappa^{(2-n)/2}\beta^{2-n}}\left[\frac{1+3\xi_nn(n-1)\kappa^{(n-2)/2}\beta^{n-2}}{\kappa+\xi_n\kappa^{n/2}\beta^n+
\frac{3}{2}\xi_n^2n^2\kappa^{(n-1)}\beta^{2(n-1)}}-\frac{1}{\kappa+\xi_n\kappa^{n/2}\beta^{n}}\right]\label{nsN}
, \end{equation}
 where the quantity $\beta$ is given by $\beta
=\frac{1}{f}\sqrt{8(1-f)(1+f[N-1])}$.

Numerically from Eq.(\ref{nsN}) we found a constraint for the
parameter $\xi_n$. Certainly, we can find the value of the
parameter $\xi_n$ giving a specific value of the parameter $f$,
when the number of e-folds $N$ and the index $n_s$ are given.
Particularly, for the values $n_s=0.967$, $N=60$  and $f=0.9$ (or
equivalently $n\simeq -0.44$), we obtained from Eq.(\ref{nsN})
that the real solution for $\xi_{n=-0.44}$ corresponds to
$\xi_{n=-0.44}\simeq-5.84M_p^{2.44}$. For the case in which
$f=2/3$ corresponds to $\xi_{n=-2}\simeq -7.42M_p^{4}$, when
$f=0.5$ the parameter $\xi_{n=-4}$ results $\xi_{n=-4}\simeq
-70.01M_p^{6}$,  and for the case where $f=0.46$ the parameter
$\xi_{n=-4.70}$ is given by
$\xi_{n=-4.70}\simeq-170.82M_p^{6.70}$.


On the other hand, the generation of tensor perturbations during
inflation would produce gravitational waves, where
its power spectrum ${\cal{P}}_{T}$, following Ref.\cite{shinji}
can be written as
\begin{equation}
{\cal{P}}_{T}\equiv A_{T}^{2}\left[\frac{k |\eta
|}{2}\right]^{3-2\nu_{T}},\label{pt}
\end{equation}
where now the tensor-amplitude  $A_{T}^{2}=
\frac{8}{F}(\frac{H}{2\pi})^{2} (\frac{1}{aH|\eta|})^{2}
[\frac{\Gamma(\nu_{T})}{\Gamma(3/2)}]^{2}$, and the quantities
$\nu_{T}$, $\gamma_{T}$ and $\delta_{T}$   are defined as
$$
\nu_{T} \equiv\sqrt{\gamma_{T} + 1/4},\,\,\,\gamma_{T}
=\frac{(1+\delta_{T})(2-\epsilon_1 +
\delta_{T})}{(1-\epsilon_1)^{2}},\,\,\,\mbox{and}\,\,\,\delta_{T}
= \frac{n\dot{\varphi}}{2H\varphi}.
$$
In this context, an important observational quantity corresponds
to the tensor to scalar ratio $r$, given by
$r={\cal{P}}_{T}/{\cal{P}}_{S}$. This ratio $r$ can be written in
terms of the slow roll parameters results\cite{noh}
\begin{equation}
r\simeq|-13.8(\epsilon_{1} +\epsilon_{3})|.\label{eq18}
\end{equation}

In this form, considering the slow parameters given by
Eq.(\ref{eq5}), we write the tensor to scalar ratio as

\begin{equation}
r\simeq\large\mid\frac{-13.8}{2}\left[C_1\frac{(1-f)}{(1-3f/2)}+C_2\frac{\varphi^{n}}{\kappa+\xi_n\varphi^n}\right]\,\varphi^{-2}\large\mid,\label{eq20}
\end{equation}
and in terms of the number of e-fold, $N$, we have
\begin{equation}
r\simeq\large\mid-\frac{13.8}{2}{\kappa\beta^2}\left[\frac{C_1(1-f)}{(1-3f/2)}
+C_2\frac{\beta^n\,\kappa^{n/2}}{\kappa+\xi_n\kappa^{n/2}\beta^n}\right]\large\mid,\label{eq29}
\end{equation}
here, we have used Eqs.(\ref{eq24}) and (\ref{eq18}).


\begin{figure}[th]
{{\hspace{-4.5cm}\includegraphics[width=4.5in,angle=0,clip=true]{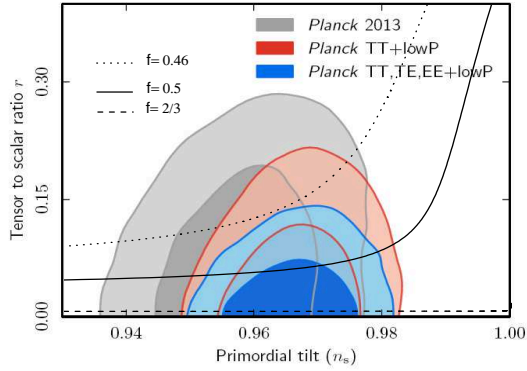}}}

{\vspace{-2.0 cm}\caption{ The plot shows the tensor to scalar
ratio $r$ versus the spectral index $n_s$. Here, from Planck data,
two dimensional marginalized constraints on the ratio $r$ and
$n_s$ (at 1 $\sigma$ confidence level, i.e., 68$\%$ and at 2
$\sigma$ confidence level, i.e., 95$\%$).
 In this  plot, the dotted, solid,
and  dashed lines correspond to the values of $f=0.46$ together
with the value $\xi_{n=-4.7}=-170,82 M_p^{6.7}$,
($f=0.5,\xi_{n=-4}=-70.01M_p^{6} $ and
($f=2/3,\xi_{n=-2}=-7.42M_p^{4} $), respectively. Here, we have
used the value $M_p=1$ .
 \label{fig1}}}
\end{figure}

In Fig.\ref{fig1}
 shows the contour plot for the tensor to scalar ratio $r$ versus
 the spectral index $n_s$, for distinct  values of the parameter
 $f$ associated to intermediate expansion of the scale factor.
 From Ref.\cite{planck2015}, we have two dimensional marginalized constraints on
 the  ratio $r_{0.002}$ and index $n_s$ (at 1 $\sigma$ confidence level i.e., 68$\%$ and at 2 $\sigma$ confidence level i.e.,
 95$\%$).
 Here, the dotted, solid,
and  dashed lines correspond to the values of $f=0.46$ together
with the value $\xi_{n=-4.7}=-170,82 M_p^{6.7}$,
($f=0.5,\xi_{n=-4}=-70.01M_p^{6} $) and
($f=2/3,\xi_{n=-2}=-7.42M_p^{4} $), respectively. In order to
write down the ratio $r$ on the spectral index $n_s$, we used
Eqs.(\ref{nsN}) and (\ref{eq29}) and we numerically obtain the
parametric plot for the relation $r=r(n_s)$. We observed that for
the values of $f\gtrsim 0.46$ (or equivalently $n\gtrsim-4.7$) and
$\xi_{n=-4.7}\gtrsim -171 M_p^{6.7}$  the model is well
corroborated by the Planck data as could be visualized from this
figure. For values of the parameter $f<0.46$ and $\xi_{n}< -171
M_p^{2-n}$ the model becomes disfavored from Planck data, because the
tensor to scalar ratio $r_{0.002}>0.1$. Also, we observed that for values
of the parameter $f\thicksim 1$, the tensor to scalar ratio $r$
tends to zero. In this way, we find that the constraints for the
parameter $f$ associated to intermediate expansion of the scale
factor is given by $0.46\lesssim f <1$ and for the parameter
$-171M_p^{2-n}\lesssim\xi_{n}<0$, from Planck data.

 On the other hand,  an analysis of all data taken by the
BICEP2 $\&$ Keck Array CMB polarization experiments together with
the Planck data, were analyzed  in Ref.\cite{BKP}. Here, combining
the  results from BICEP2 $\&$ Keck Array with the  constraints
from Planck analysis of CMB temperature, plus BAO and other data,
it was  obtained  a combined limit for the tensor to scalar ratio
$r$ at pivot scale $k_*=0.05$Mpc$^{-1}$, given by $r_{0.05} <
0.07$ at 95$\%$ (2$\sigma$ confidence level).  In this sense, the
value of $f=0.46$ together with the value $\xi_{n=-4.7}= -171
M_p^{6.7}$ (see Fig.\ref{fig1}) are disfavored from
 Ref.\cite{BKP}, since the
tensor to scalar ratio $r>0.7$.
 In this form, considering Ref.\cite{BKP} we obtain that the constraints for the
 quantity
  $f$ associated to intermediate expansion of the scale
factor is given by $0.46< f <1$ and for the parameter
$-171M_p^{2-n}<\xi_{n}<0$.

This  suggests that the function
$F(\varphi)$ can be written as
$F(\varphi)=\kappa-|\xi_n|\varphi^{-|n|}$, and assuming that the
function $F(\varphi)>0$, then the range for the scalar field
during intermediate inflation in this framework satisfies
$\varphi>(|\xi_n|/\kappa)^{(1/|n|)}$. Also, we noted that in the
limit $\varphi\rightarrow\infty$, the function $F(\varphi)$ takes
the asymptotic value
$F(\varphi)_{\varphi\rightarrow\infty}\rightarrow \kappa=(8\pi
G)^{-1}$. Thus, from the observational point of view (in
particular from the consistency relation $r=r(n_s)$), we found
that the intermediate inflation into generalized induced-gravity
scenario is less limited than the standard intermediate inflation
in which the  GR is utilized, due to the incorporation of a new
parameter, i.e., $\xi_n$.

 In the following we will mention some  constraints on the
coupling parameter $\xi_n$ obtained  in the literature, in order
to compare  with our results. In the framework of the induced
gravity inflation where the function $F(\varphi)=\xi\varphi^2$ was studied   in
Ref.\cite{kaiser}. 
   Here, for the chaotic potential case was
obtained that the constraint by the coupling $\xi_{n=2}=\xi\geq
10^{-3}$, and  for the new inflation case $\xi\leq 4\times
10^{-3}$, assuming  the constraint from scalar spectral index
$n_S$ found in \cite{kaiser}. For the case in which the function
$F(\varphi)=1+\xi\varphi^2$ with $\varphi^4$
 self-interaction, was found that $\xi\geq 4\times 10^{-3}$\cite{kaiser}.  For this same coupling function and  analyzing  the
  potentials $\varphi^p$ and  exponential,  the constraints on $\xi$ considering  the Wilkinson Microwave Anisotropy Probe (WMAP)  data
  were
  obtained in Ref.\cite{Nozari:2007eq}. Here, for the value
  $p=4$ was found that
   $\xi\leq -0.17$ and $\xi\geq 0.01$, and for the exponential potential $0.271\leq\xi\leq
   0.791$. In Ref.\cite{shinji} was analyzed the function
   $F(\varphi)=(1-\xi\kappa^2\varphi^2)/\kappa^2$ together with
   the potential $V(\varphi)\propto \varphi^p$.  The constraint obtained for the parameter $\xi$ from the $r-n_S$
   plane, for the quadratic potential ($p=2$) it was
   $\xi>-1.1\times 10^{-2}$ (2$\sigma$ bound) and  for the case in which $p=4$, 
       $\xi<-3.0\times 10^{-4}$ (2$\sigma$ bound)\cite{shinji}. In
virtue of these results, our constraint  indicates that in order
to have values of the coupling $\mid\xi_n\mid/M_p^{2-n}\,\,\,\mathcal{O}$ (1), necessary
the parameter $f$ tends to one, wherewith the tensor to scalar
ratio $r\sim 0$.

\section{Conclusions}

In this paper we have studied the intermediate inflationary model
in the context of generalized induced-gravity scenario. From a
function $F(\varphi)=\kappa+\xi_n\varphi^n$, we have found
solutions to the background dynamics under the slow-roll
approximation.  Also, we have found expressions for  the scalar
and tensor power spectrum, scalar spectral index, and
tensor-to-scalar ratio. In this sense, we have found the
constraints on  some parameters from the Planck  data. Within this
context,  we have found  from Eqs.(\ref{nsN}) and (\ref{eq29})
that the trajectories in the $r-n_s$ plane are well supported by
the data (see Fig.\ref{fig1}), and we have obtained  the
constraints on the parameters $f$ and $\xi_n$ given by
$0.46\lesssim f <1$ and $-171M_p^{2-n}\lesssim\xi_{n}<0$, respectively.
 However, considering the combined limit of Ref.\cite{BKP} in which $r_{0.05}<0.7$ at 2$\sigma$ confidence level, i.e.,
95$\%$, we have found
 the
constraints on the parameters $f$ and $\xi_n$, are given by $0.46<
f <1$ and $-171M_p^{2-n}<\xi_{n}<0$, respectively.

Also, we have noted that the intermediate inflation into
generalized induced-gravity scenario is less restricted  than the
standard intermediate inflation, due to the incorporation of a new
parameter, i.e., $\xi_n$. Thus, the inclusion of this parameter
from the function $F(\varphi)$ permits us a freedom on the
consistency relation $r=r(n_s)$.

Finally, We should mention that the effective potential given by
Eq.(\ref{eq16}) does not have a minimum. In this form,  the scalar
field does not oscillate around this
minimum\cite{RCampuzano:2005qw}, and therefore is a problem for
standard mechanism of reheating in this types of the intermediate
models\cite{Campuzano:2005qw}. We hope to return to reheating  in
the near future.

\begin{acknowledgments}
R.H. was supported by  PUENTE Grant DI-PUCV N$_0$ 123.748/2017.
\end{acknowledgments}


\end{document}